%% file: main.tex
\documentclass[conference]{IEEEtran}
\IEEEoverridecommandlockouts

\usepackage{color,xcolor}
\usepackage{comment}
\usepackage{utfsym}
\usepackage{soul}
\usepackage[ruled,vlined,linesnumbered]{algorithm2e}
\usepackage{algpseudocode}
\usepackage{array}
\usepackage{amsfonts}
\usepackage{pifont}
\usepackage{booktabs}
\usepackage{amsmath}
\usepackage{multirow}
\usepackage{threeparttable}
\usepackage{hyperref}
\usepackage{makecell}
\usepackage{graphicx}
\usepackage{subfigure}
\usepackage[para]{footmisc} 
\usepackage{balance}
\hypersetup{
  colorlinks,
  citecolor=black,
  linkcolor=black,
  urlcolor=black}
\usepackage{colortbl}
\usepackage{enumitem}
\usepackage{tikz}
\usetikzlibrary{backgrounds,matrix}
\usepackage{cleveref}
\Crefname{figure}{Fig.}{Figs.}

\usetikzlibrary{decorations.pathreplacing,calc}

\def\BibTeX{{\rm B\kern-.05em{\sc i\kern-.025em b}\kern-.08em
    T\kern-.1667em\lower.7ex\hbox{E}\kern-.125emX}}

\newcommand{\gadd}[1]{{\color{black}{#1}}}%
\newcommand{\gdel}[1]{}

\newcommand{\zdel}[1]{}

\newcommand{\ldel}[1]{}

\begin{document}

\renewcommand\ttdefault{cmtt}


\title{WISP: \underline{I}mage \underline{S}egmentation-Based \underline{W}hitespace Diagnosis for Optimal Rectilinear Floor\underline{p}lanning\vspace{-0pt}}

\author{\IEEEauthorblockN{Xiaotian Zhao, Zixuan Li, Yichen Cai and Xinfei Guo\IEEEauthorrefmark{1}}
\IEEEauthorblockA{University of Michigan – Shanghai Jiao Tong University Joint Institute \\Shanghai Jiao Tong University, Shanghai, China \\Corresponding author email: xinfei.guo@sjtu.edu.cn}
\thanks{This work has been submitted to the IEEE for possible publication. Copyright may be transferred without notice, after which this version may no longer be accessible.}
}

\maketitle

\begin{abstract}
The increasing number of rectilinear floorplans in modern chip designs presents significant challenges for traditional macro placers due to the additional complexity introduced by blocked corners. 
Particularly, the widely adopted wirelength model Half-Perimeter Wirelength (HPWL) struggles to accurately handle rectilinear boundaries, highlighting the need for additional objectives tailored to rectilinear floorplan optimization.
In this paper, we identify the necessity for whitespace diagnosis in rectilinear floorplanning, an aspect often overlooked in past research. We introduce WISP, a novel framework that analyzes and scores whitespace regions to guide placement optimization. WISP leverages image segmentation techniques for whitespace parsing, \gadd{a lightweight probabilistic model to score whitespace regions based on macro distribution,}\zdel{a Gaussian Mixture Model (GMM) for whitespace density scoring} and direction-aware macro relocation to iteratively refine macro placement, reduce \textit{wasted whitespace}, and enhance design quality. The proposed diagnostic technique also enables the reclamation of block-level unused area and its return to the top level, maximizing overall area utilization.
When compared against state-of-the-art academia placer DREAMPlace 4.1, our method achieves an average improvement of \textcolor{black}{5.4\%} in routing wirelength, with a maximum of \textcolor{black}{11.4\%} across widely-used benchmarks. This yields an average of \textcolor{black}{41.5\%} and \textcolor{black}{43.7\%} improvement in Worst Negative Slack (WNS) and Total Negative Slack (TNS), respectively. Additionally, WISP recycles an average of \textcolor{black}{16.2\%} area at the block level, contributing to more efficient top-level area distribution.
\end{abstract}


\linespread{0.75}

\section{Introduction}

\begin{figure}[t]
    \centering
    \subfigure[]{
    \includegraphics[width=1.0 \linewidth]{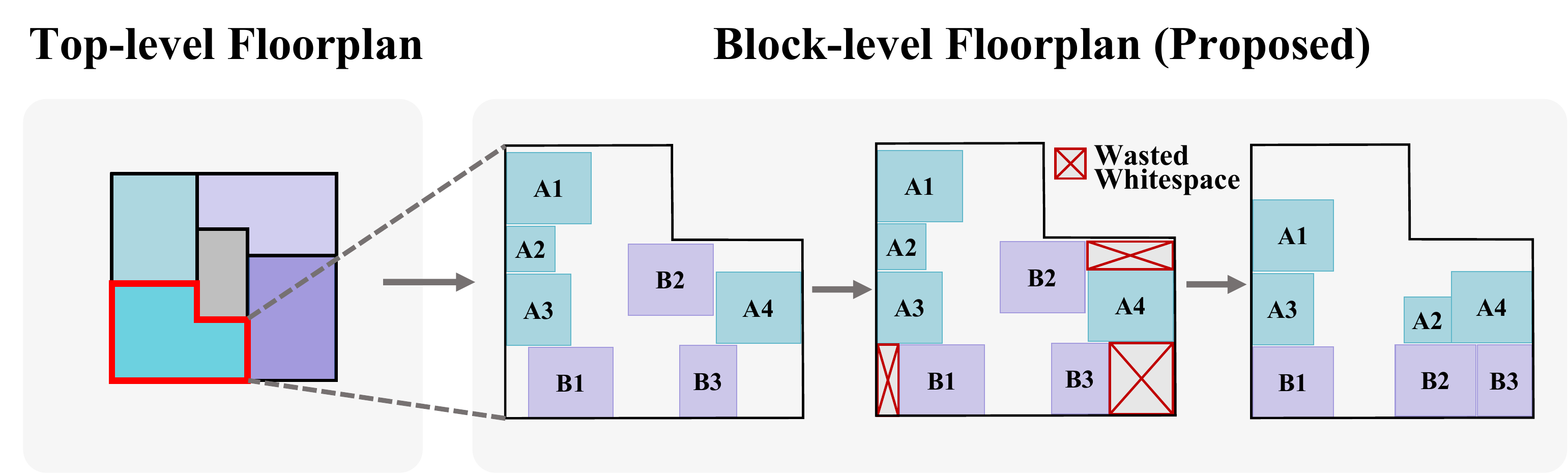}
    \label{fig1a}
    \vspace{-3pt}
    }
    \subfigure[]{
    \vspace{-3pt}
        \includegraphics[width=1.0 \linewidth]{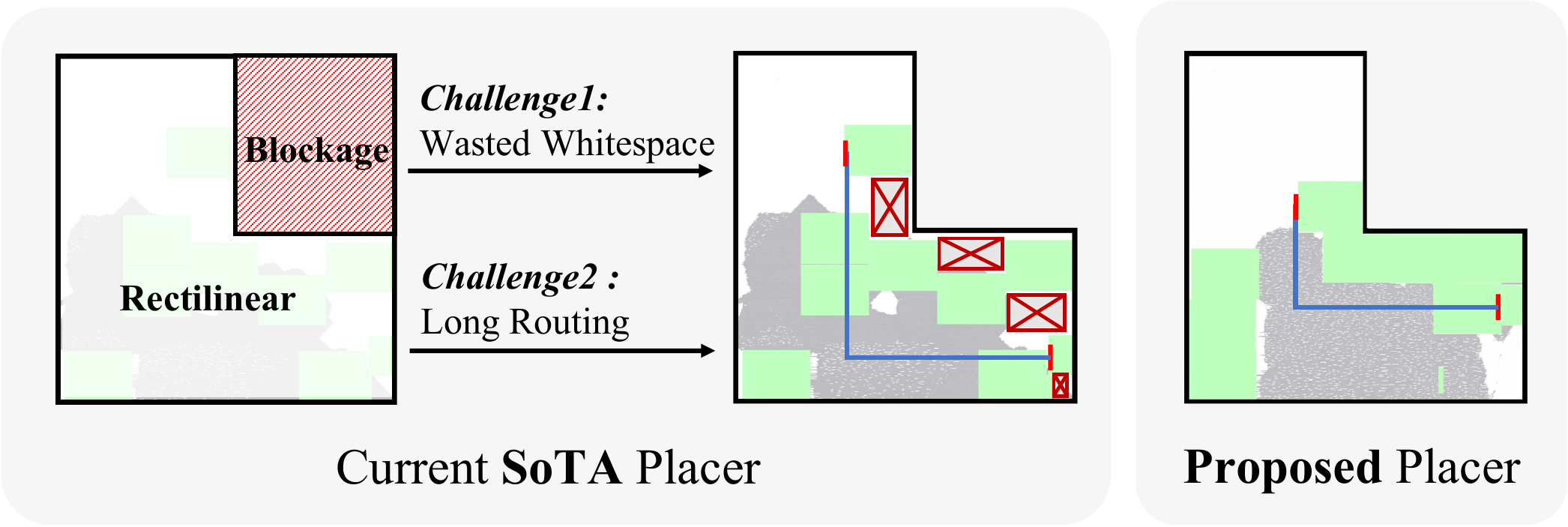}
        \label{fig1b}
        \vspace{-3pt}
    }
    \vspace{-5pt}
    \caption{(a) Illustration of the hierarchical floorplanning methodology, showcasing an optimization process from the top-level floorplan to the block-level rectilinear module floorplan. At the top level, modules are transformed from regular rectangles into rectilinear shapes to improve area utilization. The block-level floorplanning starts with a suboptimal placement, where inefficient usage of space leads to noticeable wasted whitespace. The optimized result highlights how effective whitespace analysis and reuse can significantly improve floorplan quality and area efficiency. (b) Challenges in current rectilinear floorplanning, including increased possibilities in wasted whitespace, the creation of notch-induced dead zones, and the limitations of traditional wirelength models such as HPWL in capturing the true layout characteristics of rectilinear modules.
    }
    \vspace{-10pt}
    \label{fig:fig1}
\end{figure}

With the rapid development of semiconductor technology and the exponential growth \gadd{in} chip complexity, floorplanning, \gdel{as} the first step of back-end design \gadd{process}, has \gadd{emerged as} one of the most \gadd{critical} research areas in \gadd{Electronic Design Automation (EDA)}\gdel{EDA}~\cite{ppaplacer, skyplacer}. 
\gadd{The quality of a floorplan significantly influences the overall PPA (Performance, Power, and Area) of a design, making floorplanning optimization a vital aspect of improving chip design performance.}
In modern chip design, System on Chip (SoC) has become \gadd{a} mainstream design model which relies on \gadd{numerous} internal or external IPs.  As SoCs scale in complexity and size, hierarchical floorplanning methodologies have emerged to support block reuse both within a single chip and across different chips. This approach also reduces the computational burden on placers and routers by enabling focused optimization at different hierarchy levels. Specifically, top-level floorplanning addresses the global placement of major blocks across the chip, while block-level floorplanning focuses on arranging standard cells and macros within individual blocks. As illustrated on the left side of Fig. \ref{fig1a}, to accommodate a wide range of IPs and improve area utilization, top-level floorplanning often transforms modules into rectilinear shapes, deviating from traditional rectangular shapes \cite{10137175, jigsawplanner}.

With the increasing adoption of rectilinear modules at the top-level floorplanning stage, attention naturally shifts to the block-level rectilinear floorplanning, which plays a critical role in managing the internal layout of these complex-shaped modules. However, most existing placers lack specific optimizations for rectilinear shapes. As shown in the left part of Fig. \ref{fig1b}, state-of-the-art placers typically enclose the rectilinear shape within a bounding rectangle and treat the non-functional regions, i.e., the areas outside the original shape but within the bounding box, as blockage macros. This approach introduces significant challenges compared to traditional rectangular layouts due to the irregular geometric constraints of rectilinear shapes.

The first major challenge is the increased likelihood of creating notch areas or dead placement regions, leading to inefficient utilization of space and wasted area. Secondly, the widely used Half-Perimeter Wirelength (HPWL) model, one of the key optimization metrics in traditional placer, becomes less effective in this context. Since HPWL calculates wirelength based on the bounding rectangle, it often overestimates wirelength and fails to accurately capture the true connectivity cost within a rectilinear boundary. In the absence of a wirelength model tailored to rectilinear modules, alternative approaches are needed to better address these optimization challenges.

To this end, we argue that block-level rectilinear floorplanning deserves special attention, and propose that effective whitespace analysis and utilization is key to improving its quality. As illustrated in the right part of Fig. \ref{fig1a}, the selected block represents a rectilinear module from the top-level floorplan. \gadd{Macro placement within this module creates various types of whitespace, some are fully open and suitable for standard cell insertion, while others, especially in notch areas, are less practical and wasteful. We propose a method to label and optimize these wasted whitespace regions, improving both placement quality and area efficiency. Moreover, the reclaimed whitespace can be recycled to the top-level floorplan, enabling iterative refinement of the overall chip layout.}

\renewcommand{\arraystretch}{1.7}
\begin{table}[t]
\centering
\caption{A list of recent floorplanning work.}
\vspace{-5pt}
\resizebox{\columnwidth}{!}{
\begin{tabular}{l|c|c|l}
\toprule[1.2pt]
                                       \textbf{Work} & \textbf{Year} & \textbf{Support Rectilinear?} & \textbf{Optimization for Rectilinear} \\ \midrule
\midrule[0.5pt]
\multicolumn{4}{c}{\textbf{Top-Level floorplanning}} \\  
\midrule[0.5pt]
                                        TOFU \cite{10137175} & 2023  & \checkmark                   &   Whitespace Refinement     \\
                                        JigsawPlanner \cite{jigsawplanner} & 2024  & \checkmark                 &   Jonker-Volgenant Algorithm      \\
                                        ICCAD'24 \cite{modernfloorplanning} & 2024  & \checkmark                 &  Differentiable  Mechanism       \\
\midrule[0.5pt]
\multicolumn{4}{c}{\textbf{Block-Level floorplanning}} \\  
\midrule[0.5pt]
                                       HiDaP \cite{hidap} & 2019  &  $\times$           & $\times$            \\
                                       RectilinearMP \cite{le2023toward} &  2023  &  \checkmark         &  Block region    \\
                                       Hier-RTLMP \cite{kahng2023hierrtlmp} & 2023   &  $\times$             & $\times$   \\
                                       AutoDMP \cite{autodmp}   & 2023  & $\times$          &  $\times$            \\
                                       DREAMPlace 4.1 \cite{dmp41} & 2023  & \checkmark   &    No \gadd{Specific} Optimization  \\
                                       DATE’24 \cite{10546560} & 2024  & $\times$   &    $\times$  \\
                                        IncreMP \cite{incremp} & 2024  & \checkmark              &   No \gadd{Specific} Optimization          \\\midrule
                                        \textbf{Ours}  & \textbf{2025}  & \checkmark                   &   \textbf{Whitespace
Diagnosis}       \\
                                        \bottomrule[1.2pt]
\end{tabular}
\label{tab: relatedwork}
}
\vspace{-10pt}
\end{table}

Although previous works \cite{10137175, jigsawplanner, modernfloorplanning} have explored whitespace optimization at the top-level, where many initially regular rectangular blocks are transformed into rectilinear shapes, most do not address the increasingly important challenge of rectilinear floorplanning at the block level. As the number of block-level rectilinear floorplans continues to grow, it becomes critical to develop techniques specifically targeting their optimization. As summarized in Table \ref{tab: relatedwork}, many recently proposed macro placers or floorplan optimization techniques struggle to even parse design files with rectilinear layouts. Exceptions include DREAMPlace 4.1 \cite{dmp41}, IncreMP \cite{incremp}, and RectilinearMP \cite{le2023toward}. However, the first two are not tailored for placement optimization in rectilinear floorplans, often resulting in suboptimal or unfeasible placement outcomes. RectilinearMP \cite{le2023toward} also provides optimization by using block regions to fill rectilinear spaces, transforming the floorplan into a rectangular shape and then applying simulated annealing for wirelength refinement. However, this approach still uses the HPWL model as the main optimization target, which does not address the critical aspect of rectilinear floorplan and whitespace analysis, leaving room for further improvement.


\gadd{In summary, there is a growing demand from top-level floorplanning perspectives to generate more rectilinear floorplans at the block level. 
To address this, this paper proposes a novel methodology named WISP that begins with analyzing whitespace regions using an image segmentation-inspired approach, treating the layout as an image based on initial placement. The whitespace is then scored and differentiated, enabling iterative refinement of macro placement until an optimal overall whitespace score is achieved. WISP is lightweight yet effective, fully adaptive, and specifically tailored for rectilinear floorplanning. The key contributions are summarized below.}
\begin{enumerate}
    \item \textbf{Agile Image Segmentation-based Floorplan Whitespace Parsing and Scoring}. We introduce a novel computer vision-inspired approach that leverages image segmentation to replace traditional algorithmic methods for whitespace diagnosis and parsing, significantly reducing computational complexity and improving portability. A Gaussian Mixture Model (GMM) is then leveraged to score the whitespace, enabling detailed analysis of whitespace distribution and characteristics across the floorplan [Sections \ref{subsection:CV-based Floorplan Parsing} and \ref{sec: gmm}].
    \item \textbf{Direction-Aware Macro Placement Refinement.} Guided by the whitespace score, we propose a direction-aware simulated annealing (SA) technique \gadd{that adaptively explores macro movement} for faster and more effective macro placement refinement technique [Section \ref{sec: direction}].
    \item \textbf{Quality Improvement and Potential Area Recycling.} 
    The proposed WISP flow consistently outperforms baseline placers in routing wirelength, Worst Negative Slack (WNS), Total Negative Slack (TNS), and power, within acceptable runtime. Leveraging the whitespace density scores, we introduce an area recycling strategy that reclaims low-score whitespace and feeds them back to the top-level floorplan, enabling further chip area optimization. This approach achieves an average of \textcolor{black}{\textbf{16.2\%}} area reduction over state-of-the-art mixed-size placers, while delivering superior timing performance [Section \ref{Evaluation}].
\end{enumerate}

\section{Preliminaries}
\gadd{In this section, we first discuss the current status of top-level and block-level rectilinear floorplanning methodologies. We then introduce a new term, \textit{wasted whitespace}, specifically for rectilinear floorplanning. Finally, we compare traditional whitespace extraction algorithms with our proposed image segmentation-based whitespace diagnosis approach.}

\begin{figure}[t]
    \centering
    \subfigure[Original floorplan]{
        \includegraphics[width=0.267 \linewidth]{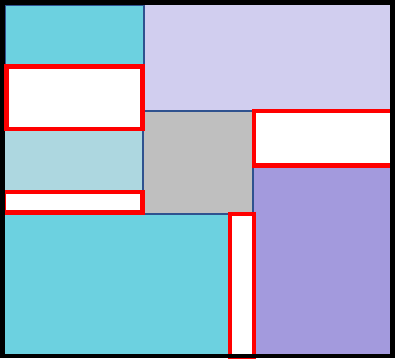}
        \label{fig2a}
    }
    \subfigure[Optimized floorplan]{
        \includegraphics[width=0.555 \linewidth]{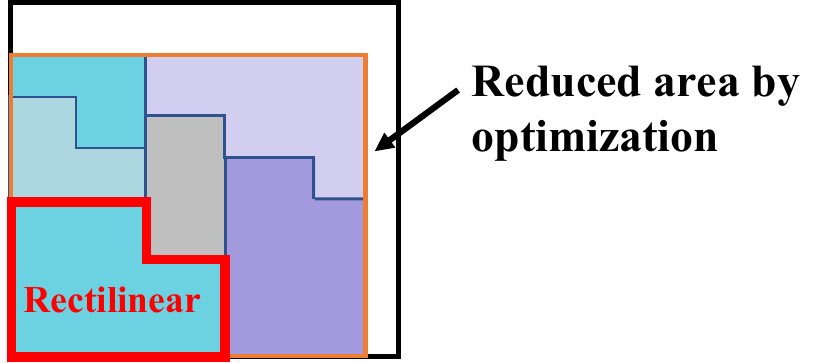}
        \label{fig2b}
    }
    \caption{\gadd{(a) Original top-level floorplan with whitespace as discussed in \cite{jigsawplanner}. (b) Top-level floorplan with modules converted into rectilinear blocks.}}
    \label{fig:top and optimized floorplan}
    \vspace{-8pt}
\end{figure}

\begin{figure}[t]
    \centering
    \subfigure[]{
    \includegraphics[width=0.455 \linewidth]{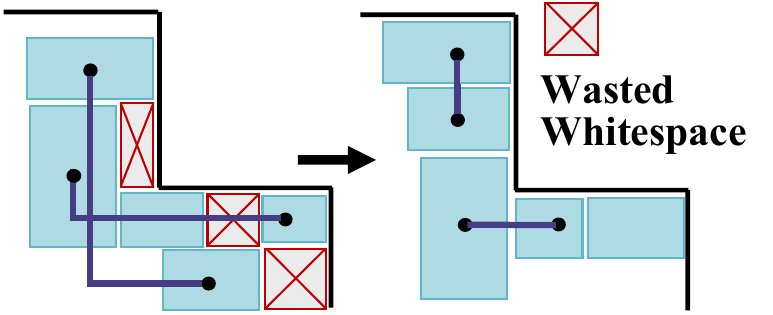}
    \label{fig3a}
    }
    \subfigure[]{
        \includegraphics[width=0.44 \linewidth]{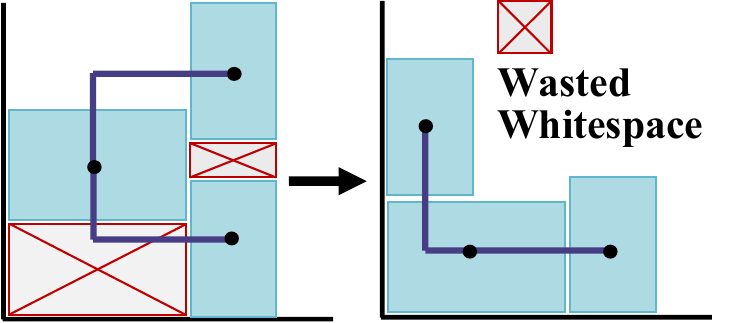}
        \label{fig3b}
    }
    \caption{\gadd{(a) \textit{Wasted whitespace} regions can lead to \gadd{long routing.} (b) Impact of \textit{wasted whitespace} in rectilinear floorplanning.}}
    \label{fig:wasted whitespace pre}
    \vspace{-8pt}
\end{figure}

\begin{figure}[t]
    \centering
    \subfigure[Traditional Algorithm]{
        \includegraphics[width=0.454 \linewidth]{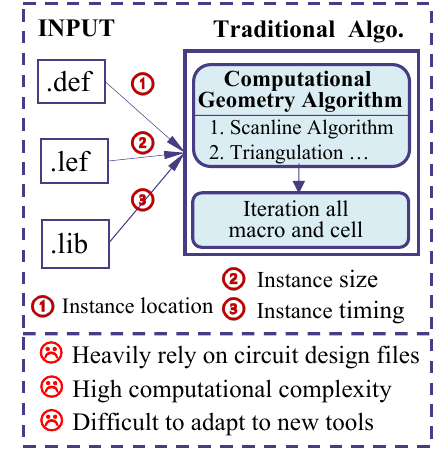}
        \label{fig4a}
    }
    \subfigure[Proposed Method]{
        \includegraphics[width=0.447 \linewidth]{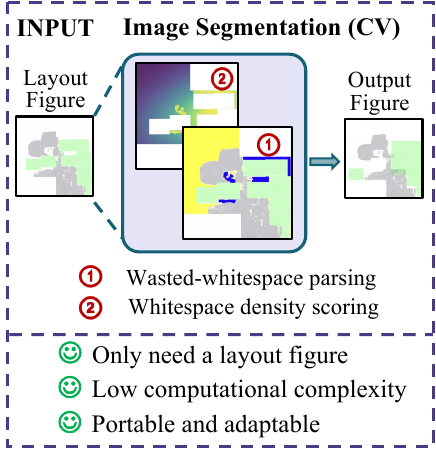}
        \label{fig4b}
    }
    \caption{Comparison between traditional algorithms and our proposed Image Segmentation-inspired method \gadd{for} whitespace analysis. 
    }
    \label{fig:compare}
        \vspace{-8pt}
\end{figure}


\subsection{Top-level Floorplanning}

\gadd{Top-level floorplanning is an initial and essential step for the whole physical design phase in a large-scale SoC. Traditional methods typically approximate modules as rectangles, as shown in Fig. \ref{fig2a}. While this simplification eases algorithmic implementation, it is no longer essential in modern designs and often limits placement flexibility—resulting in inefficient area utilization due to excessive whitespace or unavoidable overlaps. To address these limitations, recent efforts such as TOFU \cite{10137175}, JigsawPlanner \cite{jigsawplanner}, ICCAD’24 \cite{modernfloorplanning}, and FloorSet \cite{mallappa2024floorsetvlsifloorplanning} from Intel Labs have all introduced top-level floorplanning techniques that support rectilinear modules. These approaches enable more flexible and accurate modeling of complex module shapes, thereby improving overall area efficiency. As illustrated in Fig. \ref{fig2b}, converting modules into rectilinear blocks allows their irregular outlines to better conform to available whitespace, leading to a notable reduction in top-level floorplan area. These studies further underscore the importance of optimizing area utilization within rectilinear floorplans while preserving overall design quality.}
%

\subsection{\textit{Wasted Whitespace} in Rectilinear Floorplan}
\label{sec:WastedWhitespaceDefinition}
Given a layout, aside from the areas occupied by macros and cell instances, the remaining spaces are defined as \textit{whitespace regions}. \gadd{As illustrated in Fig. \ref{fig:wasted whitespace pre}, certain whitespace regions located in corner positions are relatively enclosed and blocked, making them less suitable for accommodating macros or cells. 
\textcolor{black}{Therefore, within the \textit{whitespace regions} in the floorplan, we define \textit{wasted whitespace} as small place and route (P\&R) regions that fall into a specific area threshold range (200 to 20000 pixels in our case), have relatively enclosed shapes, exhibit 0\% utilization, and can potentially cause routability issues.} As shown in the figure, \textit{wasted whitespace} contributes to increased wirelength, and minimizing it can significantly optimize wirelength \cite{incremp}.}
For rectilinear floorplans like the L-shape example in Fig. \ref{fig3a}, numerous wires might cross the \gadd{notch} regions, which are often prone to DRC violations and routing congestion. 
\gadd{Compared to rectangular floorplans, rectilinear designs with more notches are more prone to generating \textit{wasted whitespace}, which in turn introduces routing challenges. Targeting rectilinear floorplan-specific optimizations through whitespace analysis can therefore offer significant benefits in improving placement and routing efficiency.}




\begin{figure}[t]
    \centering
    \includegraphics[width=0.95\linewidth]{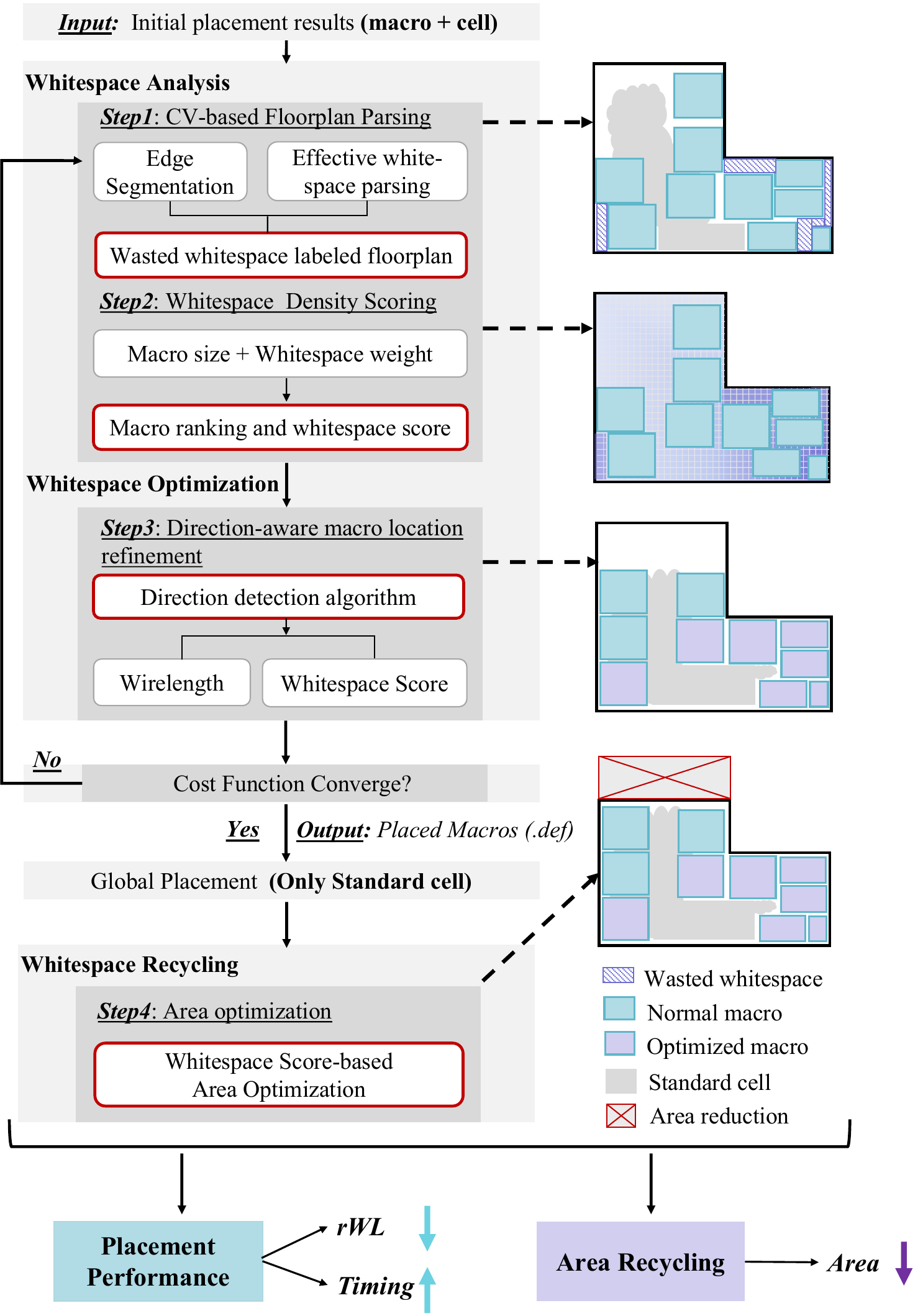}
    \caption{The overall flow of the proposed WISP methodology.}
    \label{fig:overall flow}
            \vspace{-8pt}
\end{figure}

\subsection{Traditional Algorithm vs. Image Segmentation-Inspired Parsing}

\gadd{In design practice, designers typically rely on tools to analyze spatial relationships among instances and place them closer to minimize notch regions. As shown in Fig. \ref{fig4a}, traditional algorithms, such as the classic Scanline Algorithm and Triangulation, depend on parsing circuit files (e.g., .def, .lef, and .lib) to obtain information. Due to the unique databases of different EDA tools, adapting such methods across tools and design stages is challenging. Moreover, analyzing the relative positions of all macro and cell instances requires multiple iterations, resulting in high time complexity.}
\gadd{Inspired by image segmentation in computer vision, we reformulate whitespace analysis as an image processing task, using only a layout image to analyze spatial relationships, as shown in Fig. \ref{fig4b}. This approach focuses on mask-level analysis, reducing timing complexity and offering portability across toolchains.}

\section{Proposed Methodology}

\textcolor{black}{The overall flow, shown in Fig. \ref{fig:overall flow}, begins with an initial floorplan that includes pre-placed macros and cells using any placers. First, we use an Image Segmentation-based floorplan parsing method to diagnose, parse and label the \textit{wasted whitespace} in the current floorplan. The darker purple-shaded areas in the top figure of Fig. \ref{fig:overall flow} represent the identified \textit{wasted whitespace}}. \gadd{Then, we further leverage a Gaussian Mixture Model (GMM)-based scoring mechanism to derive a global scoring map of whitespace for downstream placement optimization.} Next, a direction-aware macro placement strategy is proposed to refine the macro location. 
We select the most suitable direction based on score density for movement at each simulated annealing (SA) iteration step which ensures a more effective optimization. 
The three steps are sequentially and iteratively performed until the cost function convergence. This flow results in optimized macro placements, followed by a re-execution of global placement for standard cells to further enhance the final floorplan quality. Finally, an area optimization strategy will be implemented to achieve the area recycling that introduced in Section \ref{sec:area}. The following subsections explain detailed techniques used in each stage.

\renewcommand{\arraystretch}{1.9}
\setlength{\textfloatsep}{0pt}
\begin{algorithm}[t!]
    \small
    \setlength{\baselineskip}{1.05em}
    \caption{Image Segmentation-based Floorplan Parsing Algorithm}
    \label{alg:cv}
        \KwIn{Input floorplan image $I$ }
        \KwOut{$I$ with parsed \textit{wasted whitespace} regions }
        Convert $I$ to HSV space \\
        Generate $masks$ by edge segmentation:  \\
         $mask\_cell \gets \text{captures cell regions}$ \\
         $mask\_macro \gets \text{captures macro regions}$  \\
         $mask\_whitespace \gets \text{captures whitespace regions}$ \\
         \textcolor{blue}{\Comment{Dilate cell region to eliminate noise}}\\
         $mask\_cell_{dilated} \gets \text{cv.dilate}(mask\_cell, kernel_c, \text{iter}=3)$\\
        \ForEach{pixel $(x, y)$ in $mask\_whitespace$}{
            \If{pixel $(x, y)$ \textbf{in both} $mask\_whitespace$ and $mask\_cell_{dilated}$}{
                remove $(x, y)$ as whitespace in $mask\_whitespace$
            }
        }
        Update $mask\_whitespace$ \\
        \textcolor{blue}{\Comment{Dilate macro region to identify closed region}} \\
         $macro\_mask_{dilated} \gets \text{cv.dilate}(mask\_macro, kernel_m, \text{iter}=3)$ \\
         \textcolor{blue}{\Comment{Filter whitespace contours by area thresholds}} \\
        \ForEach {whitespace $contour$ in $mask\_whitespace$}{
            \If{overlap \textbf{exists with} $macro\_mask_{dilated}$}{
                 Calculate $contour\ area$ and skip if not in $area\ threshold$ \\
                 Label overlapping region as part of $wasted\_whitespace\_mask$ \\
                 Record \textit{wasted whitespace} pixel locations \\
            }
         Create overall $wasted\_whitespace\_mask$
        }
        \Return $I$, $wasted\_whitespace\_mask$
\end{algorithm}

\subsection{Image Segmentation-based Floorplan Parsing}
\label{subsection:CV-based Floorplan Parsing}
The complete algorithm for Image Segmentation-based floorplan parsing is listed in Algorithm \ref{alg:cv}. \textcolor{black}{The input floorplan image is mapped from design layout and scaled to a fix-sized canvas where the longest side is 800 pixels and the other side is resized based on the design aspect ratio (maximum size: 800$\times$800 pixels), which allows for a \gadd{trade-off between stable processing runtime and detailed mask extraction for any designs.}} \gadd{The process begins with the edge detection on the floorplan image using the Canny edge detection algorithm \cite{4767851}, which calculates the edge angles and labels the right-angled regions, facilitating the detection of key structural features. Based on the detected edges, three masks are generated: $mask\_cell$ for densely placed cell regions, $mask\_macro$ for macros, and $mask\_whitespace$ for whitespace areas. These steps, corresponding to lines 3-5 in Algorithm \ref{alg:cv}, establish the initial segmentation of the floorplan. 
Subsequently, since cell placement is dense yet contains gaps, these gaps are necessary and cannot be repurposed. To better extract cell mask, we introduce a morphological dilation algorithm \gadd{to fill these gaps}, as shown in Algorithm \ref{alg:cv} line 7. The dilation operation can be expressed mathematically as Equation \ref{eq: dilated} and \ref{matrix}:}
\begin{equation}
I_{\text {dilated }} = I \oplus B=\left\{(x, y) \mid(B)_{(x, y)} \cap I \neq \emptyset\right\}
\label{eq: dilated}
\end{equation}
\renewcommand*{\arraystretch}{1.1}
\begin{equation}
\label{matrix}
\resizebox{0.8\linewidth}{!}{$
I =
\begin{bmatrix}
0 & 0 & 1 & 0  \\ 
0 & 1 & 1 & 1  \\ 
1 & 1 & 1 & 1  \\
0 & 0 & 1 & 0  
\end{bmatrix} 
B = 
\begin{bmatrix}
1 & 1 \\
1 & 1   
\end{bmatrix}
I \oplus B = 
\begin{bmatrix}
1 & 1 & 1 & 1  \\
1 & 1 & 1 & 1  \\
1 & 1 & 1 & 1  \\
1 & 1 & 1 & 1    
\end{bmatrix}
$}
\end{equation}
where $\oplus$ is dilation operator, $I$ is the binary image matrix converted from HSV (Hue, Saturation, Value) image of the floorplan, $B$ is the dilation kernel matrix (a 2$\times$2 square), and $I_{\text{dilated}}$ is the result \gadd{from}\gdel{of} the dilation operation. This operation ensures that any small gaps within \gdel{the}dense regions of the floorplan are filled, \gadd{ensuring} a more precise and complete extraction of the cell mask.

\gadd{To identify the enclosed regions around macros within the whitespace, morphological dilation is employed again to extend the boundaries around macros \gadd{with the kernel matrix (a 2$\times$2 square)}, as described in Algorithm \ref{alg:cv} line 13. This action effectively pinpoints potential \textit{wasted whitespace} regions, and the regions are filtered based on an area threshold \textcolor{black}{(2000 to 20000 pixels in this work)}, as outlined in Algorithm \ref{alg:cv} line 15-20. This threshold value is determined relative to the design's area. Once the filtering is done, the remaining regions are assigned labels, resulting in a parsed floorplan with clearly marked \textit{wasted whitespace} regions.}



\subsection{Whitespace Density Scoring}
\label{sec: gmm}

After Image Segmentation-based whitespace parsing, the \textit{wasted whitespace} regions are extracted and labeled. 
\gadd{However, throughout the entire floorplan, the influence of other whitespace along with the regions with placed cells is non-uniform.}
To address this, we further analyze the entire regions excluding macros and assign a score to each pixel to generate a heatmap that visualizes the intensity of the whitespace.

\begin{figure}[t]
	\centering
	\subfigure[Parsing and Scoring]{
		\includegraphics[width=0.485\linewidth]{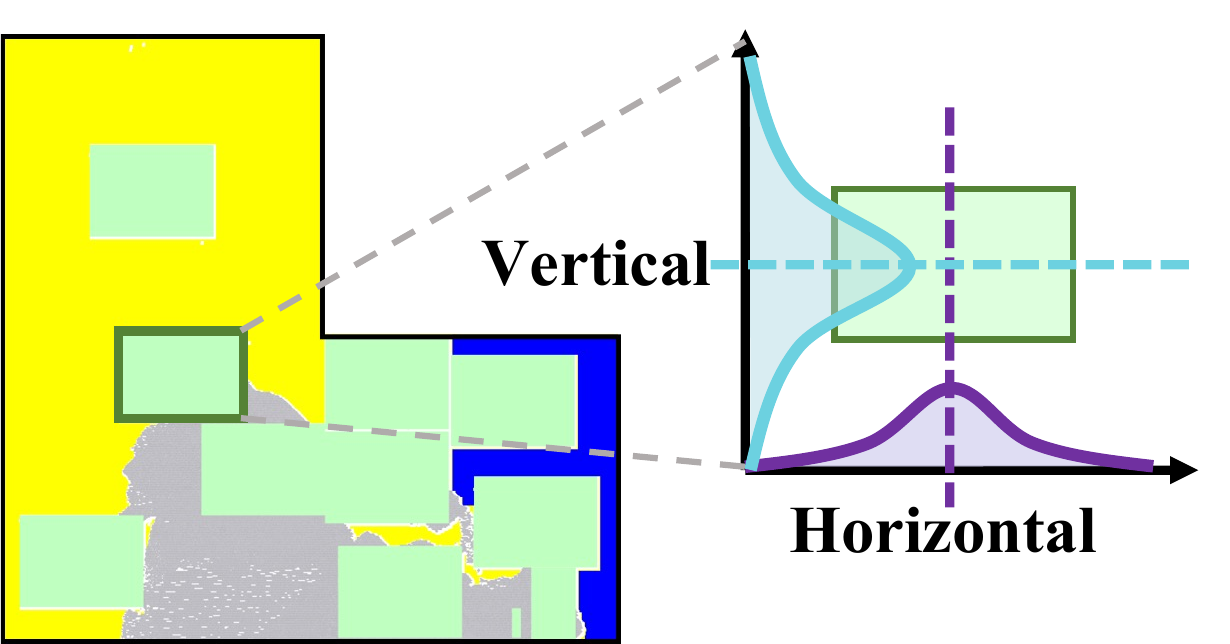}
		\label{fig:gmm_1}
	}
	\subfigure[Density heat map]{
		\includegraphics[width=0.415\linewidth]{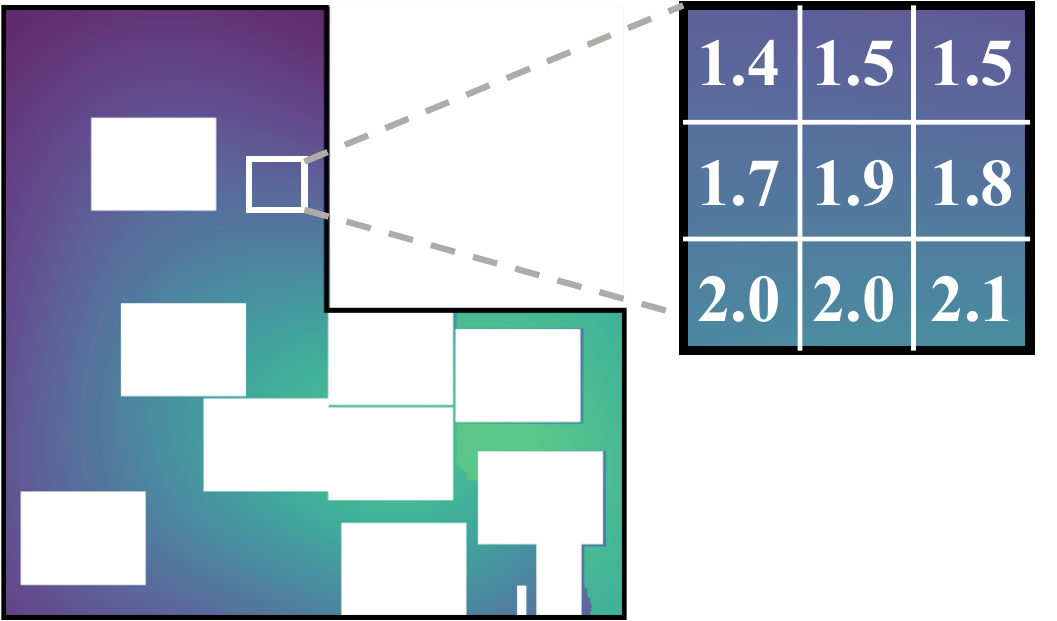}
		\label{fig:gmm_2} 
	}
    \vspace{-8pt}
\caption{(a) Illustration of scoring method on \texttt{bp\_fe} with parsed \textit{wasted whitespace} (blue region). The right part illustrates the impact of a macro on the surrounding region.  (b) The lighter-colored regions represent higher scores that require optimization.}
\label{fig:gmm} 
\end{figure}

The impact of whitespace is closely related to the spatial distribution of macros, where excessive gaps between adjacent macros degrade placement quality by reducing area efficiency and increasing wirelength \cite{incremp}. 
\gadd{In addition,} macros placed at the center of floorplan can divide the standard cells connected by the same net \gadd{into} different sub-regions, resulting in longer wirelength and more inter-region traversals.
Considering these sub-regions caused by macros, it is essential to analyze the \gadd{whitespace} between macros and their impact on the surrounding regions. 
\gadd{To address these challenges, it is crucial to capture the spatial impact of each macro which diminishes with distance.
We model the influence of each macro along both horizontal and vertical directions using Gaussian functions,} and introduce a 2-dimensional Gaussian Mixture Model (GMM) for all macros, as defined in Equation \ref{eq:gmm} and illustrated in Fig. \ref{fig:gmm_1}:
\begin{equation}
\begin{aligned}
P(s(x,y) \mid \theta) = \sum_{k=1}^{K} \alpha_k \cdot \mathcal{N}(s(x,y) \mid \boldsymbol{\mu}_k, \boldsymbol{\Sigma}_k) \\
\end{aligned}
\label{eq:gmm}
\end{equation}
where $s(x,y)$ is the pixel in 2-dimensional floorplan space, and $\mathcal{N}(s \mid \boldsymbol{\mu}_k, \boldsymbol{\Sigma}_k)$ denotes the Gaussian distribution of the $k^{th}$ macro with parameter $(\mu_k, \Sigma_k)$ and weight $\alpha_k$. 
We assume \gadd{the impact of each macro} follows a Gaussian distribution parameterized by $(\mu_k, \Sigma_k)$, as defined in Equation \ref{eq:theta_k}:
\begin{equation}
\boldsymbol{\mu}_k = 
\begin{bmatrix}
x_k \\ 
y_k
\end{bmatrix}, \quad
\boldsymbol{\Sigma}_k =
\begin{bmatrix} 
\sigma_{x}^2  \propto w & 0 \\ 
0 & \sigma_{y}^2  \propto h
\end{bmatrix}
\label{eq:theta_k}
\end{equation}
the Gaussian probability density function is calculated as:
\begin{equation}
\mathcal{N}(s \mid \boldsymbol{\mu}_k, \boldsymbol{\Sigma}_k) = 
\frac{1}{2\pi \sigma_x \sigma_y} 
e^{- \frac{1}{2} 
\left[ 
\left(\frac{x - x_k}{\sigma_x} \right)^2 + 
\left(\frac{y - y_k}{\sigma_y} \right)^2 
\right]}
\label{eq:2d_gaussian}
\end{equation}
where $\mu_k$ represents the mean of the Gaussian distribution, with $x_k$ and $y_k$ denoting the x and y coordinates of the macro, respectively.
$\sigma_{x}^2$, $\sigma_{y}^2$ are the variances in x and y dimensions, and they are proportional to the width (denoted as $w$) and height (denoted as $h$) of the macro, respectively.
The weight $\alpha_k$ for the Gaussian distribution of the $k^{th}$ macro at pixel $(x,y)$ is calculated in Equation \ref{eq:score}:
\begin{equation}
\begin{aligned}
\alpha_k = \frac{D((x,y), M_{k})}{\sum_{i=1}^K D((x,y), M_{i})} 
\end{aligned}
\label{eq:score}
\end{equation}
where $M_k = (x_k, y_k)$ represents the center coordinates of the $k^{th}$ macro, and 
$
D((x,y), M_i) = \sqrt{(x - x_i)^2 + (y - y_i)^2}
$
\gadd{denotes the Euclidean distance from the pixel $(x,y)$ to the center of the $i^{th}$ macro.}

\textcolor{black}{
Moreover, we also take the labeled \textit{wasted whitespace} into consideration. The pixel collection in the \textit{wasted whitespace} region are designated as set $A$. 
Then, we can derive the score formula for the whole floorplan outside of the macros, defined in Equation \ref{eq:score4wastedwhitespace}, where $\gamma$ is a hyper-parameter for the weight of pixel point in \textit{wasted whitespace}, we set it as $0.8$ in our process and can adjust it based on the impact intensity.
}
\begin{equation}
\resizebox{0.9\linewidth}{!}{$
P((x,y) \mid \theta) =
\begin{cases} 
\sum_{k=1}^K \alpha_k \mathcal{N}\left((x,y) \mid \theta_k\right) & (x,y) \notin A
\\ 
(1 + \gamma) \times \sum_{k=1}^K \alpha_k \mathcal{N}\left((x,y) \mid \theta_k\right) &  (x,y) \in A
\end{cases}
\label{eq:score4wastedwhitespace}
$}
\end{equation}

By calculating the value of each whitespace pixel using \gadd{Equation \ref{eq:score4wastedwhitespace}}\gdel{the $P((x,y) \mid \theta)$}, we obtain the whitespace scoring for the whole floorplan. \gadd{Taking the design \texttt{bp\_fe} as an example, its extracted \textit{wasted whitespace} and the score calculation of each macro is shown in Fig. \ref{fig:gmm_1}. The specific whitespace score of heat map is shown in Fig. \ref{fig:gmm_2}, where lighter-colored regions represent higher scores.} \gadd{It can be observed that by utilizing Gaussian-based Whitespace Scoring, the spatial relationships between a macro and its surrounding macros are effectively extracted. As illustrated in Fig. \ref{fig:gmm_2}, the region between two closely placed macros is highlighted by a lighter color, indicating a higher whitespace score and thus a higher priority for optimization. In contrast, darker-colored regions correspond to areas between macros that are farther apart, reflecting a lower interaction intensity and less need for adjustment.} This will guide the subsequent optimization of macro placement, ensuring that whitespace is utilized more efficiently while minimizing disruptions to connectivity.
 


\begin{figure}[t]
    \centering
    \includegraphics[width=0.85 \linewidth]{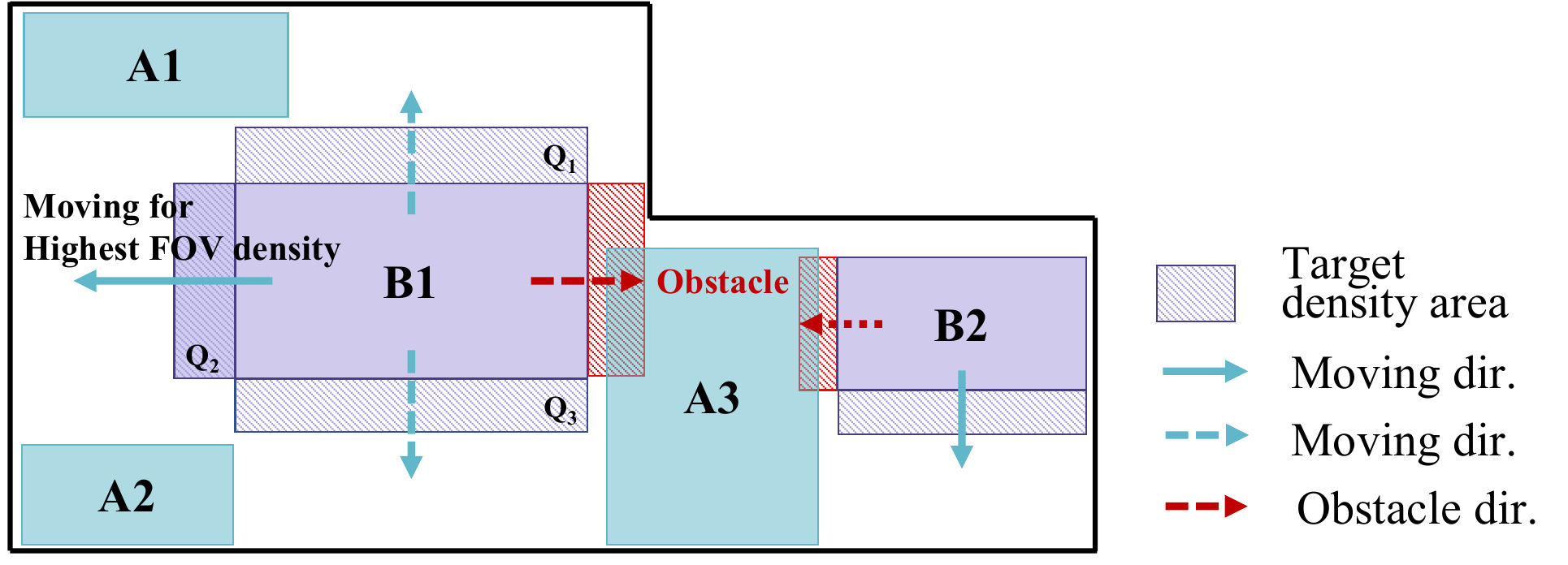}
    \caption{Direction determination in macro location optimization. By calculating the scoring density in field of vision (FOV), determine the moving direction.}
    \label{fig:dir-moving}
\end{figure}

\subsection{Direction-aware Macro Location Refinement}
\label{sec: direction}
In this step, \gadd{macro placement optimization is guided using the whitespace scoring heat map derived from the scoring calculations described earlier.} \gadd{The goal is to guide the macro movement toward regions with higher score density than its current position, effectively optimizing its placement to reduce \textit{wasted whitespace} and improve overall layout quality.} \par
\gadd{To achieve this, we propose direction-aware macro location refinement strategy, which is illustrated in Fig. \ref{fig:dir-moving}. Assuming \textit{B1} and \textit{B2} are macros to be optimized.
\gadd{To guide macro movement using the whitespace scoring heat map, we evaluate the whitespace score density within its Field of View (FOV) in four cardinal directions.} The calculation of FOV scoring density is defined in Equation \ref{eq: fov}. 
\begin{equation}
\label{eq: fov}
\begin{aligned}
FOV\ desnsity = \frac{\sum_{i=1}^{N_{FOV}} s_{i}}{N_{FOV}} 
\end{aligned}
\end{equation}
\gadd{Here, $s_i$ denotes the heat map score value at pixel $i$, and $N_{FOV}$ is the number of pixels within the FOV. This metric reflects the concentration of whitespace inefficiency in each direction. }
\textcolor{black}{The order of macro movement is based on the sum of score densities value in four FOVs, with the macro having the larger average chosen for prior movement.} If another macro exists within the FOV in a given direction, moving in that direction is avoided to prevent overlap and increased congestion. Among the remaining feasible directions, we rank the score densities and select the direction with the highest density for the macro's movement. The process is iteratively applied to all macros.} 

We then apply Simulated Annealing (SA) to refine macro positions based on a cost function that balances wirelength and whitespace distribution defined in Equation \ref{eq: cost}. \gadd{This ensures a balanced approach to improve routing efficiency while minimizing \textit{wasted whitespace}.}
\begin{equation}
\label{eq: cost}
    \textit{min}\    cost = \alpha \times Wirelength + \beta \times Whitespace\ Score
\end{equation}
Here $\alpha$ and $\beta$ are hyperparameters (set as $\alpha=0.05$ and $\beta=0.95$ in our work) and can be adjusted according to the design. \gadd{We adopt SA as our optimization strategy for macro placement due to its ability to perform local search with probabilistic exploration. Unlike global optimization methods, SA allows for localized adjustments to macro positions, suitable for adjusting macro positions based on surrounding whitespace distribution. }
\gadd{During the optimization, we set the macro movement step size to 10 pixels, allowing for effective yet controlled adjustments to macro positions at each iteration.}



\renewcommand{\arraystretch}{1.6}
\begin{table}[t!]
\centering
\caption{Benchmark Statistics.}
\label{tab:benchmarkstats}
\resizebox{1.0\linewidth}{!}{
\begin{tabular}{l|cccc}
\toprule[1pt]
\textbf{Design Name}    & \textbf{Std Cells Count} & \textbf{Macros Count} & \textbf{Macro Type} & \textbf{\# of IOs} \\ \hline \hline
\texttt{TinyRocket}    & 27217            & 2             & 1          & 269        \\
\texttt{bp\_be}         & 59882            & 10            & 3          & 3029       \\
\texttt{bp\_fe}         & 29993            & 11            & 3          & 2511       \\
\texttt{black parrot}   & 427501           & 24            & 5          & 1198       \\
\texttt{bp\_multi}      & 209086           & 26            & 6          & 1453       \\
\texttt{swerv\_wrapper} & 99750            & 28            & 3          & 1416       \\
\texttt{ariane133}     & 165953           & 133           & 1          & 495        \\
\texttt{ariane136}     &  166200                &  136             &  1          &  495       \\
\texttt{bp\_quad}      &    1238636              &  220             &    6        &  135       \\ \hline \hline
 & \multicolumn{2}{c}{\textbf{Rectilinear Num = 1}} & \multicolumn{2}{c}{\textbf{Rectilinear Num = 2}}  \\ \hline
\textbf{Floorplan Shape} & \multicolumn{2}{m{2.3cm}}{\input{Lshape}
}& \multicolumn{2}{m{2cm}}{\input{Zshape} } \\ \bottomrule [1pt]
\end{tabular}
}
\end{table}

\section{Experiment Results}
\label{Evaluation}

\subsection{Experiment Setup}
\noindent \textbf{Evaluation Flow.} The proposed methodology is implemented in python and C++, \gadd{leveraging} OpenCV library and choosing DREAMPlace 4.1 \cite{dmp41} as the baseline framework. The flow runs on Intel Core i7 11700 CPU with 128GB of memory. 
The proposed method is compared against DREAMPlace 4.1 \cite{dmp41}, a state-of-the-art open-source mixed-size placer \gadd{capable of handling rectilinear floorplans. We could not compare with \gadd{RectilinearMP \cite{le2023toward} and IncreMP \cite{incremp} due to their lack of open-source availability. Additionally, the built-in floorplanner in OpenROAD does not support rectilinear floorplanning, making direct comparison infeasible. For a fair comparison, the subsequent placement of standard cells and routing are all performed by the state-of-the-art commercial P\&R tool. All metrics are collected after post-routing optimization stage.}} \par
\noindent \textbf{Comparison Metrics.} In this study, to evaluate the improvement in routability and timing, we use post-route PPA metrics, including routed wirelength (rWL), timing (WNS and TNS), power\zdel{, since rWL alone does not fully reflect the final QoR \cite{ppaplacer}}. Additionally, in our proposed area recycling part, we record the original area and optimized area too. Runtime is also recorded to evaluate the framework’s efficiency.\par
\noindent \textbf{Benchmark Selection.} 
We select nine diverse benchmark designs from recently published work or new test suites for floorplan evaluation \cite{tilos, OpenROAD}. The synthesized netlist is obtained by running Yosys \cite{wolf2016yosys} using the open-source NanGate 45nm technology node \cite{Nangate}. \gadd{Since no publicly available rectilinear floorplans exist for benchmarking, we created in-house rectilinear floorplans. These were designed to maintain reasonable total area utilization, ensuring that baseline tools could perform initial placement without giving up, and highlight issues such as \textit{wasted whitespace} in baseline tool placement results. Table \ref{tab:benchmarkstats} summarizes the benchmark statistics. 
Floorplans with one missing corner are labeled rectilinear number of 1, and with two missing corners are labeled rectilinear number of 2.
These setups cover challenging situations where traditional placers might struggle to achieve optimal macro placement outcomes.} \par
\begin{figure}[t]
        \centering
        \includegraphics[width=0.7\linewidth]{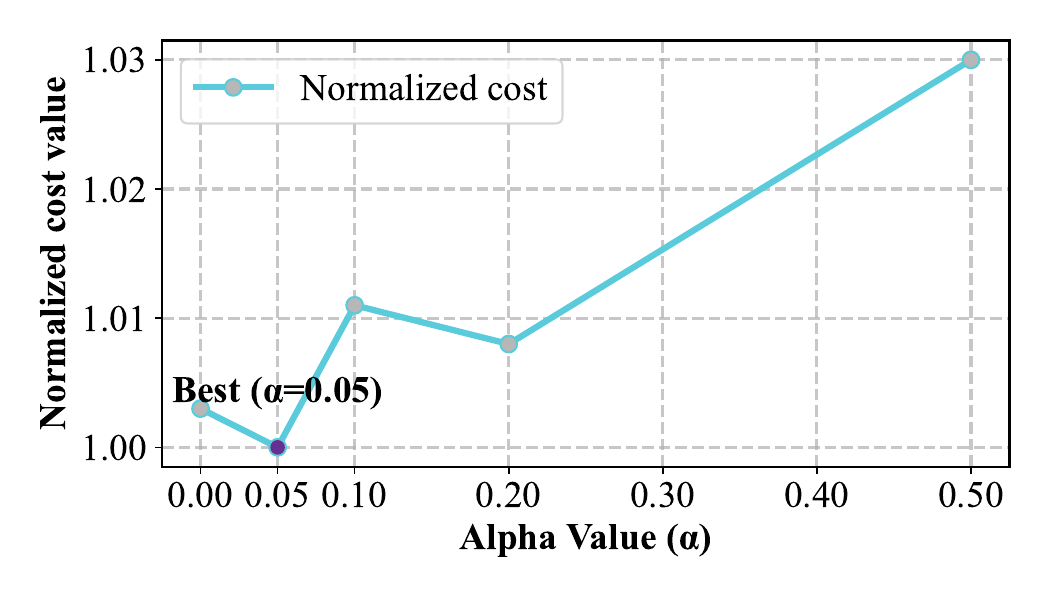}
        \vspace{-8pt}
        \caption{Hyperparameter selection of simulated annealing cost function.}
        \label{fig:sa-tune}
\end{figure}
\noindent \textbf{Hyperparameter Selection.} We determine the default values of $(\alpha, \beta)$ in our SA cost function through a sweep study. We vary $(\alpha, \beta)$ pairs such that $\alpha + \beta = 1$, run the full placement flow, and record the final cost after macro placement. The cost is normalized to the result from our default setting $(\alpha = 0.05, \beta = 0.95)$, as shown in Fig. \ref{fig:sa-tune} (on design \texttt{bp\_be}). 
This setting consistently yields competitive overall placement quality and is adopted for all designs.

\renewcommand{\arraystretch}{1.7}
\begin{table*}[t]
\caption{Comparison of Routing Wirelength (rWL), Whitespace Score, Timing and Power Results obtained after routing for 1 notch corner (Rect. Num = 1).}
\label{tab:exp_results_1}
\begin{threeparttable}
\resizebox{1.0\textwidth}{!}{
\begin{tabular}{cl|ccccc|ccccc}
\toprule [1pt]
\multicolumn{2}{c|}{\textbf{Design}}                    & \multicolumn{5}{c|}{\textbf{DREAMPlace 4.1 \cite{dmp41}}} & \multicolumn{5}{c}{\textbf{WISP (Proposed Method)}}         \\ \hline
\multicolumn{1}{l|}{\textbf{Rect. Num}} & \textbf{Name}           & \textbf{rWL (um)}  & \textbf{Score} & \makecell{\textbf{WNS (ns)}} & \textbf{TNS (ns)} & \textbf{Power (mW)} & \textbf{rWL (um)} & \textbf{Score} & \textbf{WNS (ns)} & \textbf{TNS (ns)} & \textbf{Power (mW)} \\ \midrule \midrule
\multicolumn{1}{c|}{num = 1}         & \texttt{TinyRocket}     &453557     & 174.4    &-0.032     & -0.05    & 113.3     &\textbf{444502}         &144.9  &\textbf{0.003}     & \textbf{0.00}    & \textbf{112.7}     \\
\multicolumn{1}{c|}{num = 1}         & \texttt{bp\_be}         &2547806     & 304.1      & -0.587    & -214.60     & 438.4      &\textbf{2287066}    & 287.5  & \textbf{-0.581}      & \textbf{-183.98} & \textbf{432.1}       \\
\multicolumn{1}{c|}{num = 1}         & \texttt{bp\_fe}         &1515994      & 266.3       & -0.585    &-12.70     & 267.6      &\textbf{1430214}       & 241.2 &\textbf{-0.492}     & \textbf{-10.95}    &    \textbf{265.4}    \\
\multicolumn{1}{c|}{num = 1}         & \texttt{black parrot}   &  6068799   &  281.4     &  -0.068   & -18.47    & 339.9       &  \textbf{6035678}       & 273.4       & \textbf{-0.031}       & \textbf{-1.28}    &  \textbf{335.7}  \\
\multicolumn{1}{c|}{num = 1}         & \texttt{bp\_multi}      & 3329946  & 265.6   & -0.303       & -666.51       & 378.9      &\textbf{3292197}        &  261.3 & \textbf{-0.281}     & \textbf{-585.64}     & \textbf{377.8}       \\
\multicolumn{1}{c|}{num = 1}         & \texttt{swerv\_wrapper} & 3958292    &  179.6          & -0.267        &  -401.99         & 1549.9       &  \textbf{3691505}     &  152.9    & \textbf{-0.110 }   &  \textbf{-394.50}       &\textbf{1520.2}      \\
\multicolumn{1}{c|}{num = 1}         & \texttt{ariane133 \tnote{1} }      &4791097     & 341.3      &\textbf{0.008}     & \textbf{0.00}    &  266.1     &\textbf{4639719}          & 330.8 & 0.006    & \textbf{0.00}    & \textbf{264.4}      \\ 
\multicolumn{1}{c|}{num = 1}         & \texttt{ariane136  }      &  4754798   &  377.7          &  \textbf{-0.081}       & -81.11          & \textbf{970.5 }      & \textbf{4528625}       & 345.7     & \textbf{-0.081}    & \textbf{-72.12}        & 988.5     \\ \multicolumn{1}{c|}{num = 1}         & \texttt{bp\_quad  }      & 28510588    & 574.1           &  -0.008       &     -2.34       &  \textbf{6166.9}       &  \textbf{28342822}      &   545.9   & \textbf{-0.007}    & \textbf{-2.28}        &  6187.1  \\ \midrule \hline
\multicolumn{1}{c|}{-}         & \textbf{Normalized}       & 1.041   & 1.085  & 1.486           &  1.403           & 1.005  &\textbf{1.000}       & \textbf{1.000}  & \textbf{1.000}    & \textbf{1.000}     & \textbf{1.000}       \\ \bottomrule [1pt]
\end{tabular}
}
\begin{tablenotes}
        \footnotesize
        \item [1] Use post-placement results because the placement results from DREAMPlace 4.1 fail to complete routing in commercial tool.  
\end{tablenotes}
\end{threeparttable}
\vspace{-5pt}
\end{table*}

\renewcommand{\arraystretch}{1.7}
\begin{table*}[t]
\caption{Comparison of Routing Wirelength (rWL), Whitespace Score, Timing and Power Results obtained after Routing for 2 notch corners (Rect. Num = 2).}
\label{tab:exp_results_2}
\begin{threeparttable}
\resizebox{1.0\textwidth}{!}{
\begin{tabular}{cl|ccccc|ccccc}
\toprule [1pt]
\multicolumn{2}{c|}{\textbf{Design}}                    & \multicolumn{5}{c|}{\textbf{DREAMPlace 4.1 \cite{dmp41}}} & \multicolumn{5}{c}{\textbf{WISP (Proposed Method)}}         \\ \hline
\multicolumn{1}{l|}{\textbf{Rect. Num}} & \textbf{Name}           & \textbf{rWL (um)}  & \textbf{Score} & \makecell{\textbf{WNS (ns)}} & \textbf{TNS (ns)} & \textbf{Power (mW)} & \textbf{rWL (um)} & \textbf{Score} & \textbf{WNS (ns)} & \textbf{TNS (ns)} & \textbf{Power (mW)} \\ \midrule \midrule
\multicolumn{1}{c|}{num = 2}         & \texttt{TinyRocket}     &506270     & 167.8    & -0.084   & -36.62     &  \textbf{273.3}    &  \textbf{478431}       & 159.7   &\textbf{ -0.078}    & \textbf{-31.87}    &  \textbf{273.3}   \\
\multicolumn{1}{c|}{num = 2}         & \texttt{bp\_be}          & 3442510    & 332.3    & -0.632   & -192.36     &  476.12     & \textbf{2732402}        &303.5    &\textbf{-0.628}     & \textbf{-178.48}    & \textbf{440.9}   \\
\multicolumn{1}{c|}{num = 2}         & \texttt{bp\_fe}          & 2122717    & 269.7    & -0.405   & -10.01     & 587.3      &  \textbf{1871999}       & 231.8   & \textbf{-0.061}    & \textbf{-1.45}    &\textbf{574.3}    \\
\multicolumn{1}{c|}{num = 2}         & \texttt{black parrot}     & 6870222     &293.3       & \textbf{-0.442}    & -2639.20    &381.3& \textbf{6632289}      & 268.6& -0.451    &\textbf{-2558.30}     & \textbf{380.9}        \\
\multicolumn{1}{c|}{num = 2}         & \texttt{bp\_multi}       & 3506285    & 278.3    & -0.349   & -875.98     & \textbf{382.3}      & \textbf{3422638}        &264.5    &  \textbf{-0.343}   & \textbf{-832.99}    & 383.1     \\
\multicolumn{1}{c|}{num = 2}         & \texttt{swerv\_wrapper} & \textbf{3728282}    & 157.5           &  -0.336   & -733.36      & \textbf{440.8}    & 3729285  & 133.8  &\textbf{-0.328}  & \textbf{-614.83}         &  \textbf{440.8}    \\
\multicolumn{1}{c|}{num = 2}         & \texttt{ariane133}     & 4764971    & 355.1    &\textbf{-0.088}    & -79.68     & \textbf{1006.6}      & \textbf{4590018}       & 319.7   & \textbf{-0.088}    &  \textbf{-63.62}   & 1045.8    \\ 
\multicolumn{1}{c|}{num = 2}         & \texttt{ariane136  }      &4481675    & 386.9           & -0.124        & -160.91          & 991.9      & \textbf{4315827}       & 366.1     &\textbf{ -0.073 }   & \textbf{-62.41}        & \textbf{982.3}     \\ 
\multicolumn{1}{c|}{num = 2}         & \texttt{bp\_quad}      &  29878305   &  595.8          &  -0.009       & \textbf{-3.37}          &  7177.2      & \textbf{29553113}       &   567.8   & \textbf{-0.007}    &  \textbf{-3.37}       & \textbf{7174.2} \\ \midrule \hline
\multicolumn{1}{c|}{-}         & \textbf{Normalized}        & 1.066    & 1.094    &1.343    &  1.470    &  1.008     &\textbf{1.000}       & \textbf{1.000}  & \textbf{1.000}    & \textbf{1.000}     & \textbf{1.000}       \\    \bottomrule [1pt]
\end{tabular}
}
\end{threeparttable}
    \vspace{-10pt}
\end{table*}

\begin{figure*}[!h]
    \centering
    \subfigure[\texttt{Ariane133} with Rectilinear Num = 2]{
        \includegraphics[width=0.475\linewidth]{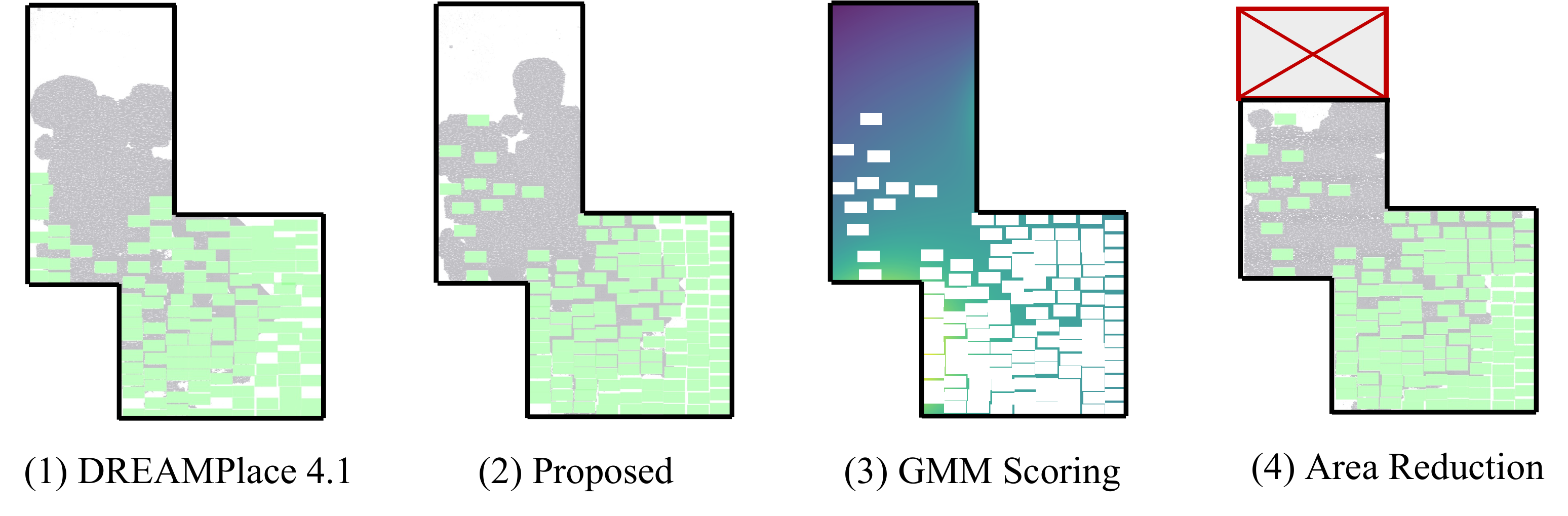}
    \label{fig:ariana} 
	}
    \subfigure[\texttt{black\_barrot} with Rectilinear Num = 2]{
        \includegraphics[width=0.475\linewidth]{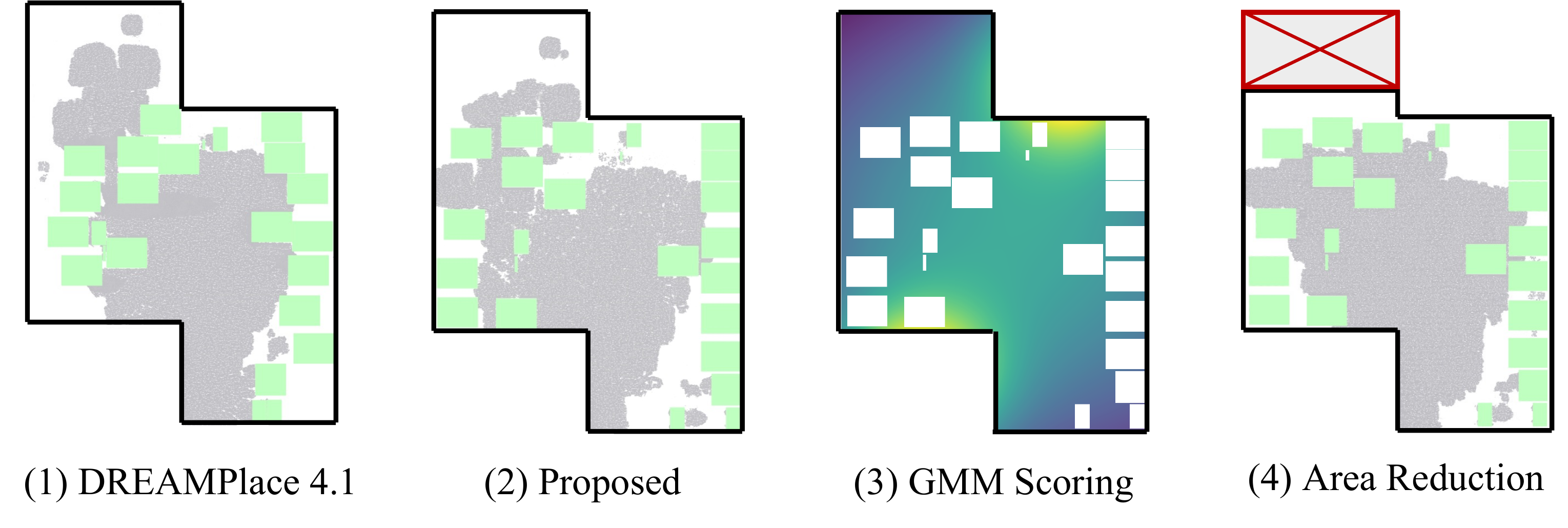}
    \label{fig:bp} 
	}
    \subfigure[\texttt{bp\_be} with Rectilinear Num = 1]{
    \includegraphics[width=0.475\linewidth]{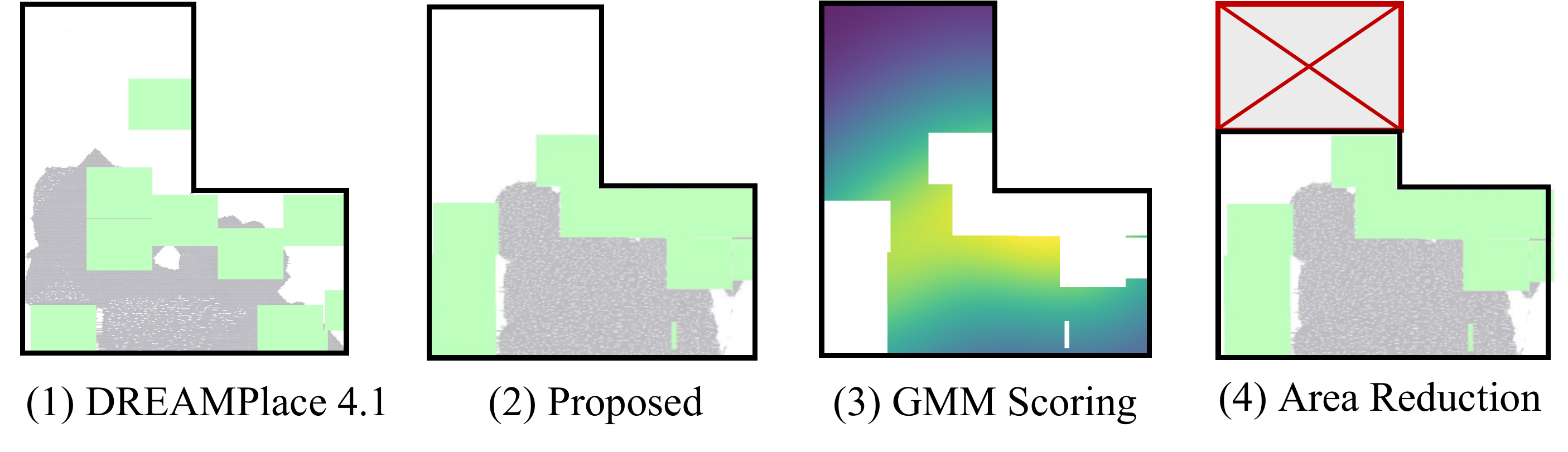}
    \label{fig:bpbe_1}
    }
    \subfigure[\texttt{bp\_be} with Rectilinear Num = 2]{
    \includegraphics[width=0.475\linewidth]{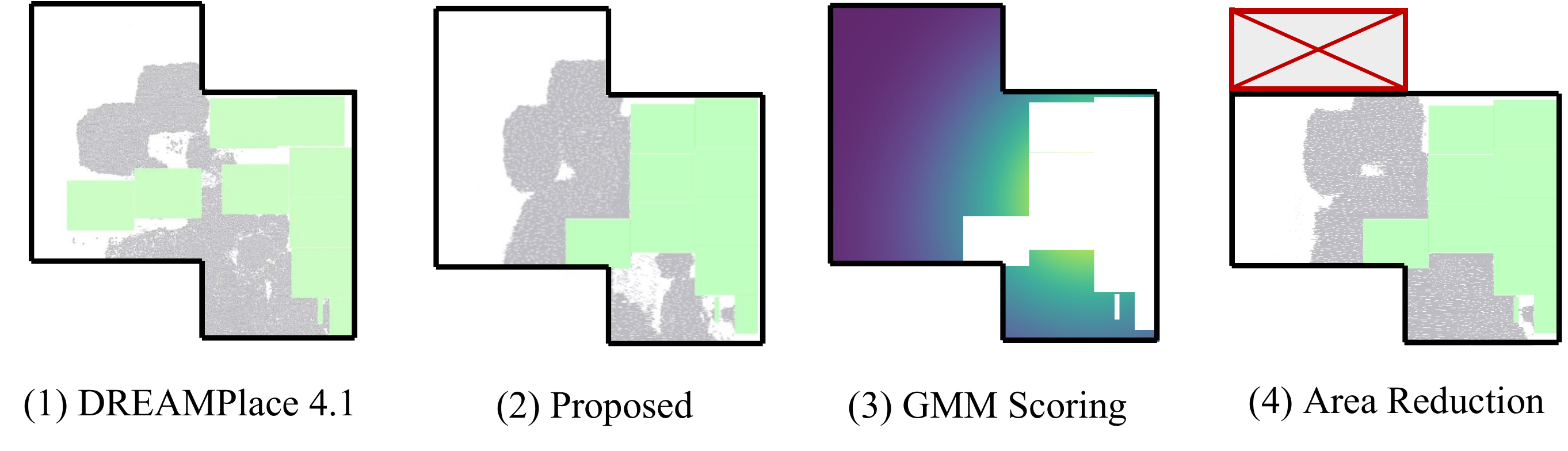}
    \label{fig:bpbe_2} 
	}
    \caption{Layouts comparison of multiple benchmark designs with rectilinear floorplan \textcolor{black}{(Rectilinear Num = 1 and Num = 2)} between DREAMPlace 4.1 \cite{dmp41} and our proposed method. A red line crossing the grey box highlights areas that can be further reduced \gadd{at the block level}.}
    \label{fig:compare bpbe}
    \vspace{-15pt}
\end{figure*} 

\begin{figure}[t]
    \centering
    \includegraphics[width=1.0\linewidth]{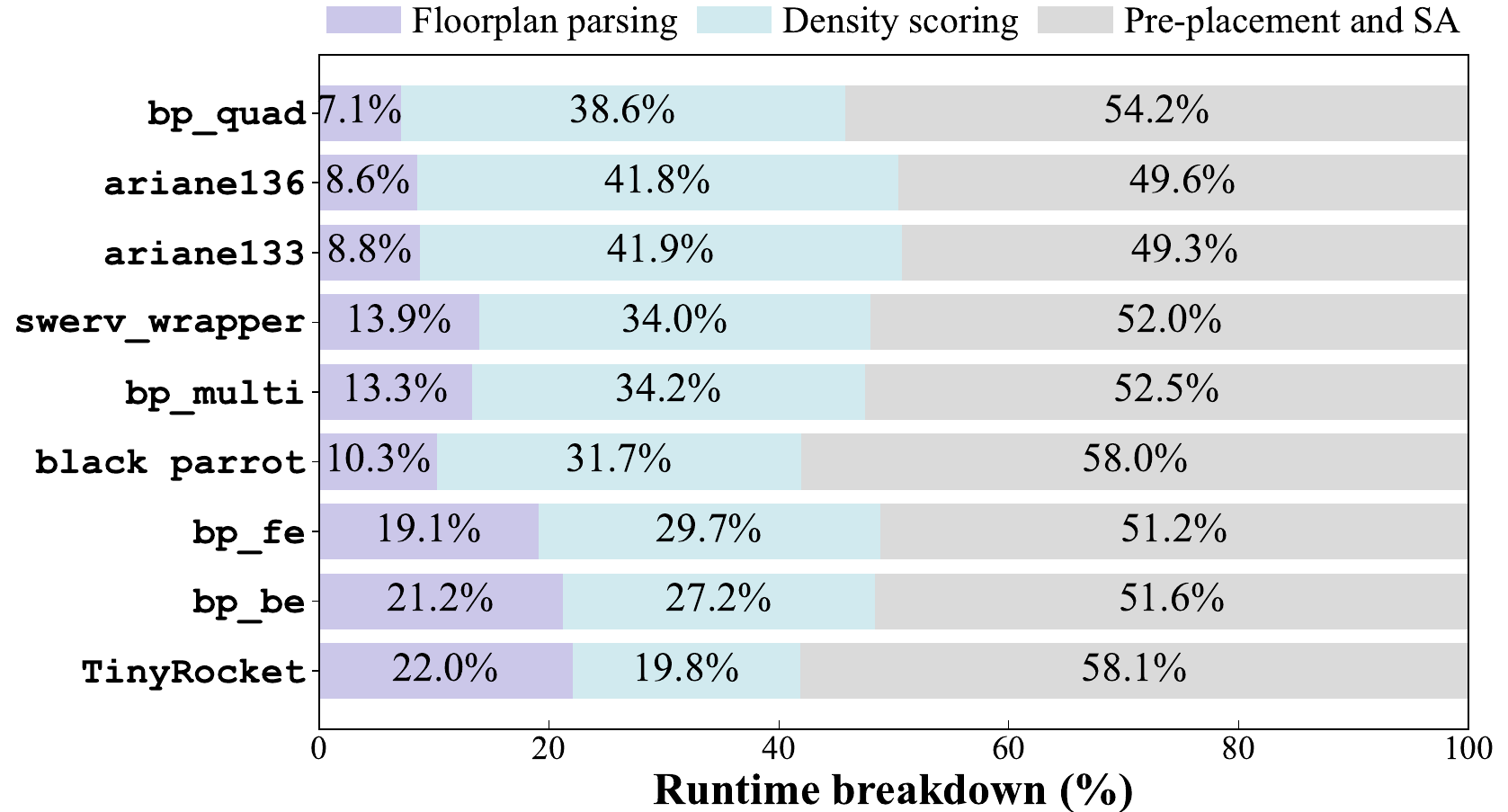}
    \caption{Runtime breakdown of the whole flow for all designs.}
    \label{fig:runtime}
    \vspace{-8pt}
\end{figure}

\renewcommand{\arraystretch}{2.1}
\begin{table}[h!]
\caption{Block-level area reduction via whitespace analysis.}
\vspace{-5pt}
\resizebox{1.0\linewidth}{!}{
\begin{tabular}{llllllll}
\toprule[1.2pt]
 & & \makecell{\textbf{Area} \\ ($um^{2}$) }  & \textbf{$\Delta$Area} & \makecell{\textbf{rWL} \\ (um)} & \makecell{\textbf{Power} \\ (mW)} & \makecell{\textbf{WNS} \\ (ns)}   & \makecell{\textbf{TNS} \\ (ns)}      \\\midrule [1.2pt]
\multicolumn{1}{l|}{\multirow{2}{*}{\begin{tabular}[c]{@{}l@{}}\texttt{Tiny Rocket}\\ \textbf{Recti Num = 2}\end{tabular}}} & \multicolumn{1}{l|}{DMP4.1\cite{dmp41}}  & 134619       & -    &  506270 &   273   &  -0.084       &  -36.62               \\ \cline{2-2}
\multicolumn{1}{l|}{}                                                                                               & \multicolumn{1}{l|}{WISP} &  \textbf{104637}        & \textbf{22.3\%}  & \textbf{471600}  & \textbf{262}  & \textbf{-0.083}      &  \textbf{-30.05}            \\ \hline
\multicolumn{1}{l|}{\multirow{2}{*}{\begin{tabular}[c]{@{}l@{}}\texttt{bp\_be }\\ \textbf{Recti Num = 1}\end{tabular}}}         & \multicolumn{1}{l|}{DMP4.1\cite{dmp41}}  & 591436 & -  &2547806  & 438   & -0.590 & -214.59 \\ \cline{2-2}
\multicolumn{1}{l|}{}                                                                                       & \multicolumn{1}{l|}{WISP} & \textbf{374434} & \textbf{36.7\% }        & \textbf{2387468}  & \textbf{432}   & \textbf{-0.580} & \textbf{-166.59} \\ \hline
\multicolumn{1}{l|}{\multirow{2}{*}{\begin{tabular}[c]{@{}l@{}}\texttt{bp\_fe}\\ \textbf{Recti Num = 2}\end{tabular}}} & \multicolumn{1}{l|}{DMP4.1\cite{dmp41}}  &  606328      &   -  &  2122716  & 587   & -0.405     &   -10.01            \\ \cline{2-2}
\multicolumn{1}{l|}{}                                                                                               & \multicolumn{1}{l|}{WISP} &  \textbf{492964}        &  \textbf{18.7\% }     & \textbf{1662803}        &  \textbf{549}       &  \textbf{-0.038}        & \textbf{-1.05}            \\ \hline
\multicolumn{1}{l|}{\multirow{2}{*}{\begin{tabular}[c]{@{}l@{}}\texttt{black\_parrot}\\ \textbf{Recti Num = 2}\end{tabular}}} & \multicolumn{1}{l|}{DMP4.1\cite{dmp41}}  & 1854920       & -     &  6870222  & 381 & -0.442      & -2639.20              \\ \cline{2-2}
\multicolumn{1}{l|}{}                                                                                               & \multicolumn{1}{l|}{WISP} &   \textbf{1680956}     &  \textbf{9.4\%} & \textbf{6625714}  & \textbf{321}  &  \textbf{-0.406}     & \textbf{-2102.50}              \\ \hline
\multicolumn{1}{l|}{\multirow{2}{*}{\begin{tabular}[c]{@{}l@{}}\texttt{bp\_multi}\\ \textbf{Recti Num = 1}\end{tabular}}} & \multicolumn{1}{l|}{DMP4.1\cite{dmp41}}   &1389160        & -    & 3506285   &\textbf{379}    &  -0.303    & -666.51       \\ \cline{2-2}
\multicolumn{1}{l|}{}                                                                                               & \multicolumn{1}{l|}{WISP} &   \textbf{1174320 }      &  \textbf{15.5\%}      & \textbf{3251848}        &  \textbf{379}       & \textbf{-0.298}         & \textbf{-584.12}            \\ \hline
\multicolumn{1}{l|}{\multirow{2}{*}{\begin{tabular}[c]{@{}l@{}}\texttt{swerv\_wrapper}\\ \textbf{Recti Num = 1}\end{tabular}}} & \multicolumn{1}{l|}{DMP4.1\cite{dmp41}}    & 1130514      & -    & 3958292   & 1549   & -0.267     & -401.99       \\ \cline{2-2}
\multicolumn{1}{l|}{}                                                                                               & \multicolumn{1}{l|}{WISP} &  \textbf{976964}        & \textbf{13.6\%}    &  \textbf{3654314}         & \textbf{1482}        &  \textbf{-0.243}        &  \textbf{-380.42}           \\ \hline
\multicolumn{1}{l|}{\multirow{2}{*}{\begin{tabular}[c]{@{}l@{}}\texttt{ariane133}\\ \textbf{Recti Num = 2}\end{tabular}}} & \multicolumn{1}{l|}{DMP4.1\cite{dmp41}}  &  1648020      & -    & 4764971   & 1006   & -0.088      &  -79.68      \\ \cline{2-2}
\multicolumn{1}{l|}{}                                                                                               & \multicolumn{1}{l|}{WISP} & \textbf{1230460}     &   \textbf{25.3\%}  &  \textbf{4613951}           & \textbf{975}    &  \textbf{-0.087}       &  \textbf{-69.51}                   \\ \hline
\multicolumn{1}{l|}{\multirow{2}{*}{\begin{tabular}[c]{@{}l@{}}\texttt{ariane136}\\ \textbf{Recti Num = 1}\end{tabular}}} & \multicolumn{1}{l|}{DMP4.1\cite{dmp41}}  & 1566414       & -    & 4754798   & 970   & \textbf{-0.081}     &  \textbf{-81.11}      \\ \cline{2-2}
\multicolumn{1}{l|}{}                                                                                               & \multicolumn{1}{l|}{WISP} & \textbf{1534407}         &    \textbf{2.4\%}    & \textbf{4600331}        &  \textbf{899}       & -0.092        &   -84.55           \\ \hline
\multicolumn{1}{l|}{\multirow{2}{*}{\begin{tabular}[c]{@{}l@{}}\texttt{bp\_quad}\\ \textbf{Recti Num = 1}\end{tabular}}} & \multicolumn{1}{l|}{DMP4.1\cite{dmp41}}  & 8817303       & -    &28510588    &  \textbf{6166}  & -0.008     & -2.34       \\ \cline{2-2}
\multicolumn{1}{l|}{}                                                                                               & \multicolumn{1}{l|}{WISP} &  \textbf{8639603}       &   \textbf{2.0\%}     & \textbf{28243888}        &   7138      &  \textbf{-0.007}        & \textbf{-1.53}            \\ \hline
\hline
\multicolumn{2}{c|}{\textbf{Average Improvement}}& \textbf{16.2\% $\uparrow$}  & - & \textbf{6.3\% $\uparrow$}  & \textbf{3.0\% $\uparrow$} & \textbf{13.9\% $\uparrow$} & \textbf{23.9\% $\uparrow$}   \\ \bottomrule[1.2pt]
\end{tabular}
}
\label{tab:casestudy}
\end{table}
\subsection{Floorplan Optimization Assessment}

Tables \ref{tab:exp_results_1} and \ref{tab:exp_results_2} summarize the routing wirelength (rWL), whitespace score, timing, and power results for designs with rectilinear numbers of 1 and 2, respectively. The results are obtained using DREAMPlace 4.1 \cite{dmp41} and our proposed method. For each design, the best result in each metric is highlighted in bold. Overall, the results demonstrate that our proposed method outperforms the DREAMPlace 4.1 in both rWL and timing.
\gadd{For designs with a rectilinear number of 1, the proposed method achieves an average improvement of 4.1\% in rWL, 48.6\% in WNS, 40.3\% in TNS, and 0.5\% in power. For rectilinear number of 2, the average improvements increase to 6.6\% in rWL, 34.3\% in WNS, 47.0\% in TNS, and 0.8\% in power.} These results indicate that the whitespace score reduction \gadd{targeted by} our method effectively translates into design performance gains \gadd{in the majority of test cases}. 
To further visualize the impact, Fig. \ref{fig:compare bpbe} shows the placement results for various benchmark designs.
As shown in Fig. \ref{fig:ariana}, \texttt{ariane133} demonstrates a significantly improved placement result. Compared to DREAMPlace 4.1, the \textit{wasted whitespace} is effectively reduced through strategic macro position adjustments, highlighting the effectiveness of our algorithm \gadd{in handling macro-rich designs}.
\gadd{A maximum rWL improvement of 11.4\% has been achieved in design \texttt{bp\_be}. The placement outcome for this design including rectilinear number of 1 and 2 are shown in Fig. \ref{fig:bpbe_1}(2) and \ref{fig:bpbe_2}(2) respectively, where it is compared against DREAMPlace 4.1 \cite{dmp41} (Fig. \ref{fig:bpbe_1}(1) and Fig. \ref{fig:bpbe_2}(1)). 
It demonstrates that our method achieves a final placement with no \textit{wasted whitespace}, significantly improving block-level area utilization.
Unusable gaps between macros are minimized, and macros are placed closer to each other and the boundary, resembling the strategy of an experienced engineer.
This not only optimizes whitespace usage but also leaves more open area for subsequent placement and routing stages.
}

\gadd{The overall timing improvement comes from the reduction of \textit{wasted whitespace} and routing wirelength (rWL). This reduction minimizes the occurrence of long and zigzag wires, which are commonly seen in notch areas and \textit{wasted whitespace} regions, thereby enhancing routability and overall timing performance of the designs.}


\subsection{Runtime Results}
\textcolor{black}{Fig. \ref{fig:runtime} provides a breakdown of the runtime for the placement and analysis stages in the proposed approach for all designs. On average, for the nine designs, the Image Segmentation-based floorplan parsing accounts for 13.81\% of the total runtime, while the GMM-based scoring contributes 33.21\%.  This highlights the relatively low complexity of analyzing, segmenting regions, and scoring whitespace pixels within the layout. In contrast, the pre-placement and SA stages dominate the runtime, collectively accounting for 52.94\%. These results underscore the computational efficiency of the parsing and scoring stages, which remain largely unaffected by design scale. However, the placement phase runtime increases significantly with design complexity, making it the primary contributor to overall runtime.}

\begin{figure}[!t]
    \centering
    \includegraphics[width=1.0\linewidth]{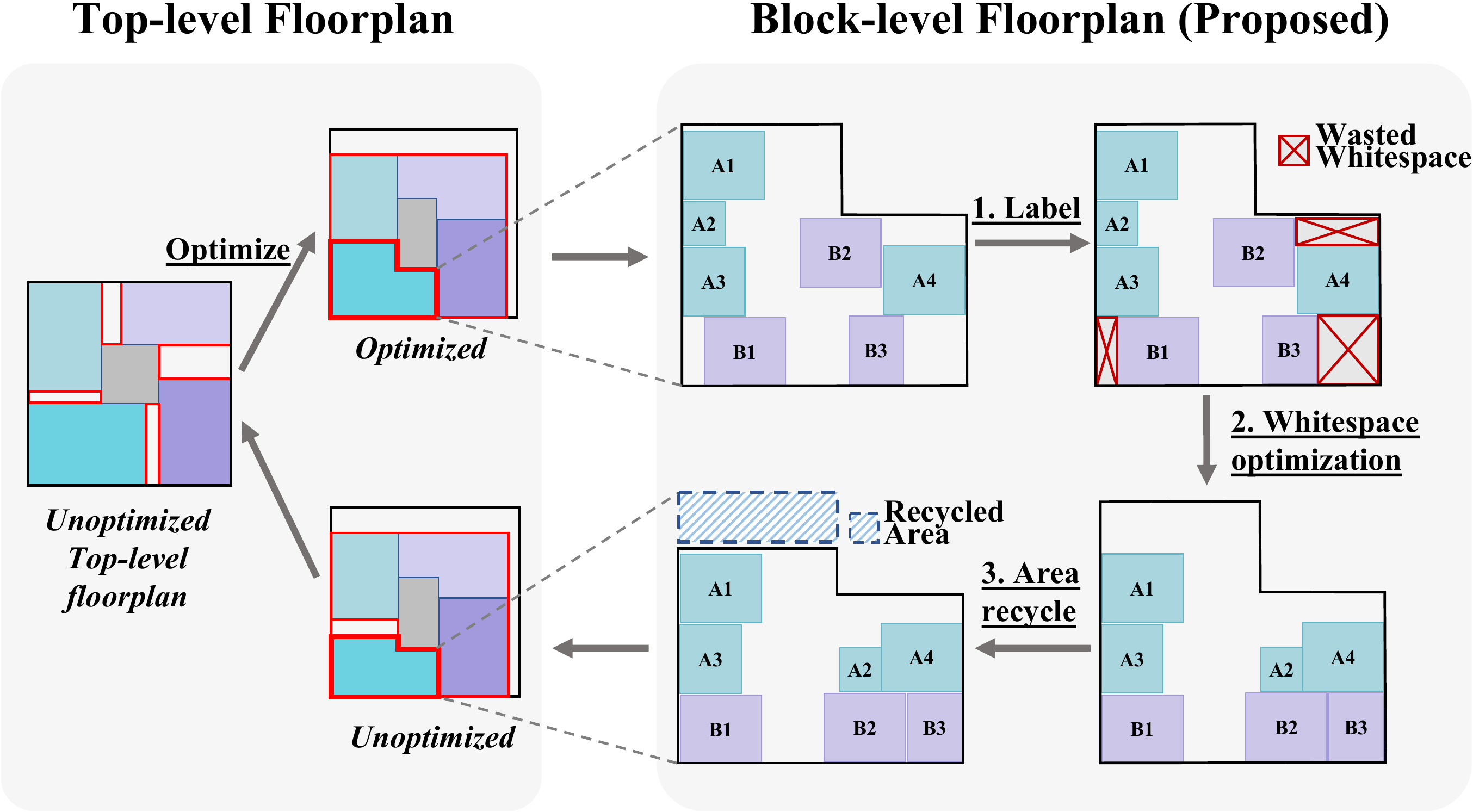}
    \caption{The illustration of an area recycling iteration from block-level rectilinear module floorplan to a top-level floorplan. }
    \label{fig:area-recycle}
\end{figure}

\subsection{\gadd{Area Recycling} —— Reclaiming \textit{Wasted Whitespace} from Block Level via Whitespace Analysis}
\label{sec:area}
The interaction between top-level and block-level owners often involves iterative negotiations to exchange area effectively. 
While recent top-level floorplanning works suggest achieving more rectilinear block-level layouts \cite{10137175,jigsawplanner,modernfloorplanning}, our whitespace analysis can identify block-level potential \textit{wasted whitespace} with lower whitespace scores. As illustrated in Fig. \ref{fig:area-recycle}, our analysis provides a systematic approach to reclaim these underutilized regions confidently and return them to the top-level, thereby maximizing overall chip area utilization. The proposed methodology enhances the baseline placer by identifying and reclaiming additional area without compromising performance or power. \gadd{These analysis results for all designs are summarized in Table \ref{tab:casestudy}, with the area reduction of 16.2\%, improvements of 6.3\% in rWL, 13.9\% in WNS and 23.9\% in TNS, which reveal the effectiveness of our method.}

This capability is also illustrated in Fig. \ref{fig:compare bpbe}, where we analyze the designs \texttt{black\_parrot} and \texttt{ariane133} with a rectilinear number of 2, as well as \texttt{bp\_be} with rectilinear numbers of 1 and 2. In Fig.~\ref{fig:ariana}(3), \ref{fig:bp}(3), \ref{fig:bpbe_1}(3), and \ref{fig:bpbe_2}(3), low-score regions are highlighted in the upper portions of the layouts, indicating substantial wasted whitespace. By removing these regions and rerunning the global placement, we obtain optimized layouts shown in Fig.~\ref{fig:ariana}(4), \ref{fig:bp}(4), \ref{fig:bpbe_1}(4), and \ref{fig:bpbe_2}(4), which demonstrate significant area reduction and improved timing. These results validate the effectiveness of our approach in reclaiming \textit{wasted whitespace} while enhancing rectilinear floorplan performance. The recycled area can be returned to the top-level floorplanner for further redistribution and overall layout refinement.

\section{Conclusions}
\gadd{In this paper, we highlight the importance of \gadd{performing} whitespace analysis in rectilinear floorplanning and propose a novel framework WISP that extracts \textit{wasted whitespace} and scores the overall whitespace distribution to guide the macro location optimization. The extracted whitespace information is seamlessly incorporated into the loss function to improve subsequent macro placement steps. 
Through extensive experiments on diverse benchmarks with different rectilinear floorplan shapes, we demonstrate that the proposed methodology significantly outperforms the leading baseline placer in routing wirelength (rWL) by 5.4\%, and in timing, with improvements of 41.5\% in WNS and 43.7\% in TNS. Additionally, WISP offers opportunity to recycle 16.2\% wasted area on average from the rectilinear blocks while maintaining or improving key design metrics. As future work, we plan to open-source this methodology, aiming to integrate it as a key feature in other open-source macro or mixed-size placers.}


\linespread{1.0}
\bibliographystyle{IEEEtran}
\balance
\bibliography{ref.bib}

\end{document}

%% file: Lshape.tex
\begin{tikzpicture}[scale = 0.2]
    \draw[thick, fill=gray!30] (0,0) -- (0,4) -- (2,4) -- (2,2) -- (4,2) -- (4,0) -- cycle;
\end{tikzpicture}

%% file: Zshape.tex
\begin{tikzpicture}[scale = 0.2]
    \draw[thick, fill=gray!30] (0,1) -- (0,4) -- (2,4) -- (2,3) -- (4,3) -- (4,0) -- (2,0) -- (2,1) -- cycle;
\end{tikzpicture}